\documentclass[twocolumn,showpacs,preprintnumbers,amsmath,amssymb,prl,superscriptaddress]{revtex4}

\usepackage{amsmath,amssymb,amsthm,color}
\usepackage{graphicx}
\usepackage{dcolumn}

\newcommand {\beq}{\begin{eqnarray}}
\newcommand {\eeq}{\end{eqnarray}}

\begin{document}

\preprint{IPMU12-0117}

\title{Intrinsic ambiguity in second order viscosity parameters \\ 
in relativistic hydrodynamics}

\author{Yu Nakayama}

\affiliation{Kavli Institute for the Physics and Mathematics of the Universe,  \\ Todai Institutes for Advanced Study,
University of Tokyo, \\ 
5-1-5 Kashiwanoha, Kashiwa, Chiba 277-8583  (Kavli IPMU, WPI), Japan}


\begin{abstract}
We show that relativistic hydrodynamics 
in Minkowski space-time has intrinsic ambiguity in second order viscosity parameters in the Landau-Lifshitz frame. This stems from the possibility of improvements of energy-momentum tensor. There exist at least two viscosity parameters which can be removed by using this ambiguity in scale invariant hydrodynamics in $(1+3)$ dimension, and seemingly non-conformal hydrodynamic theories can be hiddenly conformal invariant.
\end{abstract}

\maketitle

\section{1. Introduction}
Relativistic hydrodynamics has attained its renewed interest in the context of gauge/gravity correspondence, where some viscosity parameters of strongly coupled gauge theory can be predicted from holography. Recent experimental results in high-energy/high-luminosity collisions of hadrons seem to confirm the success of the gauge/gravity correspondence.

The relativistic hydrodynamics can be understood as an effective field theory with systematic derivative expansions. The relativistic Navier-Stokes equation is nothing but the conservation of the energy-momentum tensor, which any low energy physics must obey, and therefore, it is universal as long as we are concerned about the near-equilibrium slow excitations (compared with the characteristic scale of the microscopic theory).

The structure of the first order space-time derivative corrections to the ideal fluid is long known (see e.g. \cite{LL}). The derivative corrections induce the viscosity, and makes the problem non-linear and dispersive.
There, the choice of the frame, intimately related to the field redefinition ambiguity, is emphasized to fix what we mean by the velocity vector. The convenient frame vastly used in the literature is the so-called Landau-Lifshitz frame, which also plays a central role in this paper.

At the second order in space-time derivative expansions, the structure of the viscosity tensor has been less known. It is only in the recent literatures \cite{Baier:2007ix}\cite{Bhattacharyya:2008jc}\cite{Romatschke:2009kr}\cite{Bhattacharyya:2012ex}\cite{Banerjee:2012iz}, the systematic understanding has been pursued (see also \cite{IS}\cite{Koide:2006ef}\cite{Denicol:2010xn} for a related problem of causality issues in relativistic hydrodynamics). In this paper, we would like to point out the intrinsic ambiguity in second order viscosity parameters in the Landau-Lifshitz frame.

The ambiguity  stems from the possibility of improvements in energy-momentum tensor. In relativistic field theories, the energy-momentum tensor is not uniquely specified, and it is determined up to the improvement term whose conservation is
 automatic and hence does not affect the global generator. While the improvement term must play a trivial role in relativistic hydrodynamics, the Landau-Lifshitz condition forces us to redefine our fields since the improvement does change the value of the energy-momentum tensor at each space-time point. In this way, the appearance of the hydrodynamic equation can be changed by the improvement.
We find that there exist at least two viscosity parameters which can be removed by using this ambiguity in scale invariant hydrodynamics in $(1+3)$ dimension. In particular, seemingly non-conformal hydrodynamic theories can be hiddenly conformal invariant due to this ambiguity.

In this paper, we focus on the relativistic hydrodynamics in $(1+3)$ dimension in Minkowiki space-time. We comment on the gravitational effects in section 5. The discussions are almost parallel in other dimensions, but in $(1+1)$ dimension, the structure may be a little special so we summarized some key features in Appendix A.

\section{2. Improvements of energy-momentum tensor and frame choice}
In relativistic field theories, the energy-momentum tensor in Minkowski space-time is symmetric and conserved from the Poincar\'e invariance. However, its form is intrinsically ambiguous. One can always ``improve" the energy-momentum tensor by adding the term
\begin{align}
\delta T_{\mu\nu} = \partial^\rho \partial^\sigma L_{\mu\rho \nu \sigma} \ ,
\end{align}
where $L_{\mu\rho\nu\sigma}$ has the symmetry of the Riemann (or Weyl) tensor.
The added term is automatically symmetric and conserved: $\partial^\mu \delta T_{\mu\nu} = \partial^\mu \delta T_{\nu\mu} = 0$ due to the symmetry without using any equations of motion. The improvement does not change the generators of the Poincar\'e group.
In the context of the general relativity, the ambiguity comes from the curvature coupling $\int d^4 x \sqrt{g} R_{\mu\rho\nu\sigma} L^{\mu\rho\nu\sigma}$. 

For the discussions of the conformal field theories, the most relevant improvement is the symmetric tensor $L_{\mu\nu}$ improvement (we use mostly plus convention for the Minkowski metric $\eta_{\mu\nu}$)
\begin{align}
\delta T_{\mu\nu} &=  \frac{1}{2}(\partial_\mu \partial_\rho L^{\rho}_{\ \nu} + \partial_\nu \partial_\rho L^{\rho}_{\ \mu} - \partial^2 L_{\mu\nu} -\eta_{\mu\nu} \partial_\rho \partial_\sigma L^{\rho\sigma}) \cr
&+  \frac{1}{6}(\eta_{\mu\nu} \partial^2 L^{\rho}_{\ \rho} - \partial_\mu\partial_\nu L^{\rho}_{ \ \rho}) \ , \label{pimp}
\end{align}
which changes the trace of the energy-moment tensor as
\begin{align}
\delta T^{\mu}_{\ \mu} = \partial^{\mu}\partial^{\nu} L_{\mu\nu} \ .
\end{align}
From this expression, one can conclude that the necessary and sufficient condition  that  scale invariant  field theories are conformal invariant is (by assuming the existence of the conformal current) that the trace of the energy-momentum tensor is written as $T^{\mu}_{\ \mu} = \partial^{\mu}\partial^{\nu} L_{\mu\nu}$ (see e.g. \cite{Polchinski:1987dy}\cite{Nakayama:2010zz}).

In relativistic hydrodynamics, the fundamental equation, i.e. relativistic Navier-Stokes  equation is nothing but the energy-momentum conservation, so it seems that the improvement has nothing to add to the physical content of the hydrodynamics. While this argument is completely true, the appearance of the equation can be different due to the improvement. This is what we would like to call the ambiguity in relativistic hydrodynamics. 
The relativistic hydrodynamics is formulated as systematic space-time derivative expansions of the effective theories of fluid. The ambiguity starts to appear  at the second order of the expansions because the improvement term we reviewed is the second order in space-time derivative. 

The seemingly innocuous improvement term in energy-momentum tensor changes the appearance of the hydrodynamic equation due to the choice of the frame. Within the systematic space-time derivative expansions, the energy-momentum tensor in relativistic hydrodynamics takes the following form
\begin{align}
T_{\mu\nu} = \epsilon u_\mu u_\nu + p P_{\mu\nu} + \tau_{\mu\nu} \ ,
\end{align}
where $u^\mu$ is the normalized velocity vector (i.e. $u^\mu u_\mu = -1$), and $P_{\mu\nu} = \eta_{\mu\nu} + u_\mu u_\nu$. The higher derivative terms are all included in the viscosity tensor $\tau_{\mu\nu}$. As argued in \cite{LL}, however, in order to talk about the viscosity, we have to make more precise what we mean by the velocity vector because we can enjoy freedom of field redefinition. So-called Landau-Lifshitz frame used heavily in the literature demands that $u^\mu \tau_{\mu\nu} = 0$.  The crucial point for our discussion is that this condition depends on the value of the energy-momentum tensor at each space-time point, so the definition of $u_\mu$ in the Landau-Lifshitz frame does depend on the improvement of the energy-momentum tensor.

This means that once we try to impose the Landau-Lifshitz condition on $u_\mu$ by using the improved energy-momentum tensor, we have to redefine $u_\mu$ at the second order in space-time derivative expansions. Then we have to substitute it back into the relativistic Navier-Stokes equation or conservation equation of the energy-momentum tensor. This apparently changes the form of the equation while the physics remains the same. 

Suppose we add the improvement term $\mathcal{T}_{\mu\nu}$  of order $\mathcal{O}(\partial^2)$ to the energy-momentum tensor. In order to satisfy the Landau-Lifshitz frame condition, we have to redefine $\tilde{\epsilon} = \epsilon + \delta\epsilon $, and $\tilde{u}_\mu = u_\mu + \delta u_\mu$ (and therefore $p$ due to the equation of states), so that $ \tilde{u}^\mu \tilde{\tau}_{\mu \nu} = 0$.

In the approximation we are interested in (i.e. second order in space-time derivatives), we have
\begin{align}
\delta \tau_{\mu\nu} &= \mathcal{T}_{\mu\nu} + \delta \epsilon u_\mu u_\nu + \epsilon \delta u_\mu u_\nu + \epsilon u_\mu \delta u_\nu \cr
&+ \delta p P_{\mu\nu} +  p \delta u_\mu u_\nu + p u_\mu \delta u_\nu 
\end{align}
and the Landau-Lifshitz condition becomes
\begin{align}
u^\mu \mathcal{T}_{\mu \nu} = \delta \epsilon u_\nu + \epsilon \delta u_\nu + p \delta u_\nu
\end{align}
by using $u^\mu \delta u_\mu = 0$. 
The necessary transformation is, therefore,
\begin{align}
\delta \epsilon &= -u^\mu u^\nu \mathcal{T}_{\mu\nu} \cr
\delta u_\mu &= \frac{1}{\epsilon + p} \left( u^\rho u^\sigma \mathcal{T}_{\rho\sigma} u_\mu + u^\nu \mathcal{T}_{\nu \mu} \right) \ . \label{trans}
\end{align}

In this way, the relativistic hydrodynamic equation changes its form by the field redefinition induced by the Landau-Lifshitz condition. We will discuss its implication in the next section. We emphasize, however, that the physical contents of the relativistic hydrodynamics cannot be changed by the field redefinition. Thus, the change of the appearance of the equation only indicates the intrinsic ambiguity of the relativistic hydrodynamic equations of motion beyond the second order in space-time derivatives.

\section{3. Ambiguity in second order viscosity parameters}
The viscosity tensor can be systematically expanded in the Landau-Lifshitz frame with respect to space-time derivative as
\begin{align}
\tau_{\mu\nu} = \tau^{(1)}_{\mu\nu}(\partial) + \tau^{(2)}_{\mu\nu}(\partial^2) + \cdots \ .
\end{align}
The first order term is
\begin{align}
\tau^{(1)}_{\mu\nu} = -\eta \sigma_{\mu\nu} - \zeta P_{\mu\nu} \Theta \ .
\end{align}
Here, $\eta$ is the so-called shear viscosity and $\sigma_{\mu\nu}$ is the shear tensor defined as
\begin{align}
\sigma^{\mu\nu} = P^{\mu\sigma} P^{\nu \rho}\left(\frac{\partial_\sigma u_\rho + \partial_\rho u_\sigma}{2} - \frac{\Theta}{3} \eta_{\sigma \rho} \right) \ ,
\end{align}
where $\Theta = \partial^\mu u_\mu$. $\zeta$ is the so-called bulk viscosity. Since the ambiguity that we have discussed in the last section only affects the form of the energy-momentum tensor beyond the second order in space-time derivatives, the  viscosity parameters $\eta$ and $\zeta$ are free from the ambiguity associated with the improvement, and the both are physical.

The second order term can be expanded as
\begin{align}
\tau^{(2)}_{\mu\nu}& = \tau (D\sigma)_{\langle \mu\nu\rangle} + \lambda_0 \Theta \sigma_{\mu\nu} + \lambda_1 \sigma_{\langle \mu }^{\ \rho}\sigma_{\rho \nu \rangle} \cr 
& + \lambda_2 \sigma_{\langle \mu }^{\ \rho} \omega _{\rho \nu \rangle} +  \lambda_3 \omega_{\langle \mu }^{\ \rho} \omega _{\rho \nu \rangle} + \lambda_4 a_{\langle \mu}a_{\nu \rangle} \cr
&+ P_{\mu\nu}\left( \zeta_1 D \Theta + \xi_1 \Theta^2 + \xi_2 \sigma^2 + \xi_3 \omega^2 + \xi_4 a_\mu^2 \right) \  . \label{svis} 
\end{align}
Here we have introduced the derivative $D=u^\mu \partial_\mu$ in ``time" direction. $\omega_{\mu\nu}$ is the vorticity tensor
\begin{align}
\omega^{\mu\nu} = P^{\mu\sigma} P^{\nu \rho}\left(\frac{\partial_\sigma u_\rho - \partial_\rho u_\sigma}{2}\right) \ ,
\end{align}
and $a_\mu$ is the acceleration vector
\begin{align}
a_\mu = D u_\mu \ .
\end{align}
The braket in the tensor indicates the traceless symmetrization with the projection orthogonal to $u_\mu$ direction:
\begin{align}
X_{\langle \mu \nu \rangle} = P_\mu^{\rho}P_\nu^{\sigma} \left(\frac{X_{\rho\sigma}+ X_{\sigma\rho}}{2} -\frac{X_{\alpha \beta} P^{\alpha \beta}}{3} g_{\rho \sigma} \right) \ .
\end{align}
Finally, $\sigma^2 = \sigma_{\mu\nu} \sigma^{\mu\nu}$ and $\omega^2 = \omega_{\mu\nu} \omega^{\mu\nu}$. 

In these expressions, we have used the basis used in \cite{Bhattacharyya:2012ex}. Since one can use the equations of motion
\begin{align}
-D \epsilon - (\epsilon + p) \Theta = \mathcal{O} (\partial^2) \cr
(\epsilon + p) D u_\mu + P_{\mu\nu} \partial^\nu p = \mathcal{O} (\partial^2) \ , \label{leom}
\end{align}
at each order of the space-time derivative expansions,
it is possible to choose a different basis for the viscosity tensor by using derivatives on $\epsilon$ and $p$ instead (see e.g. \cite{Baier:2007ix}\cite{Romatschke:2009kr} where different basis is used).

We now discuss the effects of improvement and subsequent field redefinition to preserve the Landau-Lifshitz condition. By assuming locality and the absence of non-trivial dynamical degrees of freedom other than the uncharged fluid, the relevant improvement terms are given by
\begin{align}
L_{\mu\nu} = L_1(T) \eta_{\mu\nu} + L_2(T) u_\mu u_\nu \ , \label{improve}
\end{align}
where $L_1(T)$ and $L_2(T)$ are arbitrary functions of the local temperature $T$. Note that there is no candidate for $L_{\mu\nu\rho\sigma}$  constructed out of $T$ and $u_\mu$ which cannot be reduced to $L_{\mu\nu}$.\footnote{For simplicity, we have assumed parity in \eqref{svis}, but within the approximation we are interested in, the parity non-invariant term does not appear in the improvement.}

Correspondingly, the necessary field redefinition is obtained by substituting \eqref{improve} into \eqref{pimp} and \eqref{trans}. The final expression is lengthy, but let us show the contribution from $L_1(T)$:
\begin{align}
\delta \epsilon &= \frac{1}{3}(\partial^2 + u^\mu u^\nu \partial_\mu \partial_\nu)L_1(T) \cr
\delta u_\mu &= \frac{1}{3(\epsilon+p)} [ u_\mu(-\partial^2 - u^\rho u^\sigma \partial_\rho \partial_\sigma)L_1(T)  \cr
& +(u_\mu \partial^2 - u^\rho \partial_\rho \partial_\mu)L_1(t) ] \ ,
\end{align}
which are subject to the lower order equations of motion \eqref{leom} to relate $\partial_\mu T$ with derivatives on the velocity vector.
This gives at least two functional degrees of freedom $L_1(T)$ and $L_2(T)$ that contribute arbitrarily to the second order viscosity parameters. Since the detailed form how it contributes to the viscosity parameters depend on the details of the equations of states $\epsilon(T)$ and $p(T)$, we study one simple but practical situation for illustration.

Suppose we start with the conformal fluid, where $\epsilon = 3p = T^4$. The original energy-momentum tensor must be traceless so that $\zeta = \zeta_1 = \xi_1 = \xi_2 = \xi_3 = \xi_4 = 0$. We attempt to improve (or rather un-improve) the energy-momentum tensor by
\begin{align}
L_{\mu\nu} =l_1 T^2 \eta_{\mu\nu} + l_2 T^2 u_\mu u_\nu \ .
\end{align}
The power of $T$ is chosen so that the manifest scale invariance is intact. Now, in the new Landau-Lifshitz frame with the improvement, the trace of the energy-momentum tensor is non-zero. Indeed, the transformation gives the trace part of the second order viscosity tensor as
\begin{align}
\delta \tau^{(2)}_{\mu\nu} =& \frac{P_{\mu\nu}}{3}\left( l_1 T^2 (-\frac{4}{3}D\Theta -\frac{4}{9}\Theta^2 -2\sigma^2 + 2\omega^2 +4 a_\mu^2) \right. \cr
&+ \left. l_2 T^2 (\frac{4}{3}D\Theta +\frac{4}{9}\Theta^2  + 
\sigma^2 - \omega^2 - 2 a^2_\mu) \right) \cr
&+ \mathrm{traceless \ terms} 
\end{align}
Therefore, at least two in viscosity parameters of scale invariant hydrodynamics are unphysical, and one can remove it by choosing a suitable improvement term in energy-momentum tensor. We also note that if these two parameters are the only source of the breaking of the conformal invariance, we can happily improve it so that the hidden conformal invariance is recovered. 

To end this section, we would like to mention the effect of the improvement and the field redefinition on the entropy current. While the precise form of the entropy current with the second order corrections in the space-time derivative expansions is not completely determined in the literature (see e.g. \cite{Loganayagam:2008is}\cite{Bhattacharyya:2012ex}), we can still predict the intrinsic ambiguity from our consideration. In conformal fluid, zeroth order entropy current is given by
\begin{align}
J_\mu = \epsilon^{3/4} u_\mu \ .
\end{align}
Thus, the second order ambiguity associated with the improvement must be
\begin{align}
\delta J_\mu =& \frac{3}{4} \epsilon^{-1/4} u_\mu \delta \epsilon + \epsilon^{3/4} \delta u_\mu  \cr
=& \frac{\epsilon^{-1/4}}{4}[(u_\mu\partial^2 -u^\rho  \partial_\rho \partial_\mu) L_1(T)]  \cr
=&\frac{\epsilon^{-1/4}}{2}l_1( u_\mu(2a_\rho^2 + \omega^2 -\sigma^2 -D\Theta) \cr
&- \Theta a_\mu + D^2 u_\mu)   \  
\end{align}
subject to the equation of motion \eqref{leom}, which is used in the third line. Here, we have again neglected the tensor improvement contribution by setting $L_2(T) = 0$ for simplicity.

\section{4. Comment on Locality}
In microscopic theories without scalar fields, it is typical that there are no candidates for the improvement of the energy-momentum tensor in terms of local microscopic fields. This does not invalidate our discussions because the macroscopic fields that we are dealing with are space-time averaged fields whose microscopic origin may be strongly coupled. For instance, there is no field such as temperature $T$ in microscopic theory.
Of course, when the microscopic theory does possess the microscopic ambiguity in  energy-momentum tensor (e.g. $\xi R \phi^2$ term for scalar fields), our ambiguity has the microscopic origin. 

However, this may lead us to the question what kind of improvement we should allow. If we are free from any microscopic constraint, is it possible to use whatever tensor $L_{\mu\nu\rho\sigma}$ to improve the energy momentum tensor? For instance, is it allowed to remove the trace part by solving $\partial^2 L = T^{\mu}_{\mu}$ as $L = \partial^{-2} T^{\mu}_{\ \mu}$?
Such a non-local improvement would not violate any conservation law, and the physical prediction would remain the same {\it if} the equation of motion we use were exact. However, we use the space-time derivative expansions for our systematic approximation scheme, so for its self-consistency, we have to restrict ourselves to the local improvement of energy-momentum tensor. 

This has led us to the proposed improvement only by using $u_\mu$, $T$ and their derivatives. Otherwise, the systematic space-time derivative approximation is totally messed up. From the physical viewpoint, it is satisfactory because the non-local field redefinition would apparently violate, or make obscure some important physical requirements such as causality and unitarity.

If we have more conserved currents than the energy-momentum tensor (e.g. baryon number), we can introduce other improvement terms involving the chemical potential and so on. The generalization is obvious.

\section{5. Discussions}
In this paper, we have discussed the intrinsic ambiguity in relativistic hydrodynamics with the second order viscosity parameters in Minkowski spacetime. The ambiguity comes from the possibility of improvement in energy-momentum tensor and the subsequent field redefinitions in order to preserve the Landau-Lifshitz frame condition. We conclude that at least two viscosity parameters are unphysical. 

Since energy-momentum tensor itself cannot be a local observable without specifying which energy-momentum tensor one uses, the ambiguity we discussed suggests that there is no physical way to measure the second order viscosity parameters in experiments without fixing the ambiguity (in the Landau-Lifshitz frame). 
It is of great importance to discover an ambiguity-free way to measure the viscosity parameters in generic situations since the structure depends on the equations of states and they are quite involved in the current approach. Maybe the Landau-Lifshitz frame is not a good frame beyond the second order space-time derivative expansions.

The discussion in this paper is entirely macroscopic. In the future works, we would like to study the microscopic formulation to understand the ambiguity better by using the generalization of the linear response theory such as Kubo formula. Again, when we talk about the microscopic field theory, the energy-momentum tensor is ambiguous, and the correlation functions are ambiguous, too. The generic structure of the vacuum correlation functions of the energy-momentum tensor in $(1+3)$ dimension without conformal invariance is not yet known (let alone in the finite temperature case), so theoretically it is an interesting open problem.\footnote{We would like to stress that the discussion in \cite{Baier:2007ix}\cite{Bhattacharyya:2008jc} fixes the ambiguity that we discussed by choosing the traceless conformally improved energy-momentum tensor. This is a natural choice for conformal field theories.}

The constraint on the second order viscosity parameters have been discussed in the literature \cite{Romatschke:2009kr}\cite{Bhattacharyya:2012ex}\cite{Banerjee:2012iz}, and it is interesting to see whether the intrinsic ambiguity we found are consistent with their constraint. Since we do not know the microscopic origins of these constraint, it is again important to understand the structure of the correlations functions of energy-momentum tensor. Note that some of the crucial steps in their discussions rely on putting the theory on a curved space-time. Some of the constraints therefore have less practical values because there is no practical way to measure the viscosity parameters with space-time curvatures in table-top (i.e. hadron collider) experiments (exceptions would be fluid dynamics inside stellar objects like neutron stars).

In this paper, we have not turned on any background gravity. After turning on the gravity, however, the ambiguity we have discussed becomes physical because the coupling to the gravity does pick up one specific energy-momentum tensor, and therefore, the improvement starts to have physical difference. While this does not invalidate our study relevant for table-top experiments where we cannot see any gravitational effects at all, and we do not know a-priori which energy-momentum tensor couples to gravity, it is theoretically interesting to see what will happen with the additional gravitational viscosity parameters and to establish how the gravitational interaction resolves the ambiguity we have discussed. 
 
In the Einstein gravity, the conservation of the energy-momentum tensor is necessary for its consistency due to the Bianchi identity. It, however, means that one can always introduce ``gravitational energy-momentum tensor" $R_{\mu\nu} - \frac{R}{2} g_{\mu\nu}$ without spoiling the conservation (at the sacrifice of renormalizing the Newton constant). Again the added term is trivially conserved, but if we impose the Landau-Lifshitz  condition, we have to redefine the hydrodynamic fields, and it induces the curvature viscosity terms (see e.g. \cite{Romatschke:2009kr}\cite{Bhattacharyya:2012ex}). Without dynamical gravity (e.g. in AdS/CFT), we cannot avoid the ambiguity.

Finally, it would be interesting to address these questions in the context of gauge/gravity correspondence or hydrodynamics/gravity correspondence. 

\section*{Acknowledgments}
The author thanks  Jyotirmoy Bhattacharya for discussions.
The work is supported by the 
World Premier International Research Center Initiative of MEXT of
Japan. 

\appendix

\section{Appendix: In $(1+1)$ dimension}
The discussion in the most part of the paper can be readily applicable in any dimensions with a few modifications in coefficients. In $(1+1)$ space-time dimension, however, the structure of the improvement is slightly different due to the fact that all the curvature invariants are expressed by Ricci scalar only. Thus, in $(1+1)$ dimension, there is no tensor improvement available, but only the scalar improvement 
\begin{align}
\delta T_{\mu\nu} = (\partial_\mu \partial_\nu - \eta_{\mu\nu} \partial^2) L(T)
\end{align}
is effective. 

In $(1+1)$ dimension, there is no traceless viscosity tensor in the Landau-Lifshitz frame, so we have only three second order viscosity parameters
\begin{align}
\tau^{(2)}_{\mu\nu} = P_{\mu\nu} \left( \zeta_1 D \Theta + \xi_1 \Theta^2 + \xi_4 a_\mu^2 \right) \ ,
\end{align}
where $\omega_{\mu\nu} = \sigma_{\mu\nu} = 0$ identically. They are not traceless, so it must be (improved to be) zero for conformal fluid.

Generically, we have one functional ambiguity in second order viscosity parameters, but let me focus on the ambiguity in the conformal fluid, where the situation becomes singular due to its special kinematics.
By repeating the similar analysis in section 3, now with $L(T) = \log T$, we can show that
\begin{align}
\delta{\tau}^{(2)}_{\mu\nu} = 0 \ 
\end{align}
after going back to the Landau-Lifshitz frame 
because $\partial^2 \log T = 0$ for conformal ideal fluid. 
Therefore, in the particular case of the conformal fluid in $(1+1)$ dimension, there is no ambiguity in the second order viscosity parameters. We stress that this is peculiar to the kinematics of $(1+1)$ dimension with conformal invariance.

The corresponding ambiguity in the entropy current, whose zeroth order term is $J^\mu = \epsilon^{1/2} u^\mu$, is given by
\begin{align}
\delta J_\mu &= \frac{1}{2} \epsilon^{-1/2} u_\mu \delta \epsilon + \epsilon^{1/2} \delta u_\mu  \cr
&= \frac{\epsilon^{-1/2}}{2}[(-u^\rho  \partial_\rho \partial_\mu) \log T]  \cr
&= \frac{\epsilon^{-1/2}}{2}( D^2 u_\mu -\Theta a_\mu - (D\Theta) u_\mu) \ ,
\end{align}
where we have used the equations of motion. Although it looks non-trivial,  its divergence vanishes identically (up on the usage of the equation of motion) as expected.

In $(1+1)$  dimensional relativistic scale invariant field theories, the microscopic argument \cite{Polchinski:1987dy} shows that they are actually conformal invariant.  It would be interesting to see how this can be obtained from the macroscopic relativistic hydrodynamics. The non-decreasing properties of entropy can be derived from Boltzmann's H-theorem, whose microscopic foundation is just unitarity. We believe studies of the entropy current must give a clue to this problem, but we will leave  it for future studies.









\end{document}